\documentclass[preprint,12pt,review]{elsarticle}

\newcommand{\cretin}{\textit{Cretin}}
\newcommand{\hydra}{\textit{Hydra}}
\usepackage{subcaption}
\usepackage{svg}
\usepackage{multirow}
\usepackage{amsmath}
\usepackage{booktabs}
\usepackage{float}
\usepackage{color}\definecolor{AggieMaroon}{cmyk}{.15,1.00,.39,.69}
\definecolor{MaizeIsAnArrogantShadeOfYellow}{cmyk}{0.0,.18,1.00,0.0}

\journal{XXXXXXXX}

\begin{document}

\begin{frontmatter}

\title{Neural Network Surrogate Models for Absorptivity and Emissivity Spectra of Multiple Elements}

\author[inst1]{Michael D. Vander Wal}
\ead{mvander5@nd.edu}

\author[inst1]{Ryan G. McClarren}
\ead{rmcclarr@nd.edu}

\author[inst2]{Kelli D. Humbird}
\ead{humbird1@llnl.gov}

\affiliation[inst1]{%
  \organization={University of Notre Dame - Department of Aerospace and Mechanical Engineering},
  \streetaddress={University of Notre Dame},
  \city={Notre Dame},
  \state={Indiana},
  \country={USA},
  \postcode={46556}}

\affiliation[inst2]{%
  \organization={Lawrence Livermore National Laboratory},
  \streetaddress={7000 East Ave},
  \city={Livermore},
  \state={California},
  \country={USA}}

\begin{abstract}
  Simulations of high energy density physics are expensive in terms of computational resources. In particular, the computation of opacities of plasmas in the non-local thermal equilibrium (NLTE) regime can consume as much as 90\% of the total computational time of radiation hydrodynamics simulations for high energy density physics applications. Previous work has demonstrated that a combination of fully-connected autoencoders and a deep jointly-informed neural network (DJINN) can successfully replace the standard NLTE calculations for the opacity of krypton. This work expands this idea to combining multiple elements into a single surrogate model with the focus here being on the autoencoder.
\end{abstract}


\end{frontmatter}

\maketitle

\section{Introduction}
Nuclear fusion has been called the "Holy Grail"\cite{gilliam1975monty} of energy \cite{Claessens2020} due to the seemingly limitless amount of energy that controlled nuclear fusion could make available. Over the preceding decades there has been much investment into various different methods to try achieve this milestone. One of these methods is inertial confinement fusion (ICF). The National Ignition Facility (NIF) at Lawrence Livermore National Laboratory (LLNL) is exploring inertial confinement fusion. An integrated ICF experiment with deuterium and tritium filled fuel capsules is quite expensive, on the order of \$1+ million dollars per experiment. To utilize resources efficiently, ICF researchers rely heavily on simulation as an integral part of the experiment design and selection process.

ICF experiments and simulations at NIF involve the firing of many lasers at a small target, often with the goal of trying to break even between the energy put into the fuel capsule and the energy liberated via nuclear fusion. The ultimate goal is to get closer to ignition --  the point at which the energy liberated via nuclear fusion surpasses the energy of the laser drive. In these experiments, the lasers heat a hollow, cylinder-like object made of gold called a hohlraum, producing an X-ray radiation drive. Inside the hohlraum there is also a microcapsule of the fuel, often deuterium-tritium ice encased in an ablator material such as beryllium or high-density carbon \cite{kline}. The capsule is compressed to extreme density, pressure, and temperature via the X-ray drive, reaching conditions in which the deuterium and tritium begin to fuse, producing high energy neutrons and alpha particles.

In simulations of ICF experiments, atomic physics are of particular importance. The atomic physics account for the various electronic transitions of atoms of the various elements involved. These electronic transitions and ionizations are important because they are what give rise to the spectral lines and broadband radiation that characterize the absorptivities and emissivities of these elements. These are in turn necessary for energy transfer calculations.

The accounting for the electron transitions leads to the usage of either local thermal equilibrium (LTE) or non-local thermal equilibrium (NLTE) physics. That is, it is largely an accounting for how many atoms have an electron that moves up or down in the atoms' energy levels as well as electrons that become free electrons or captured electrons \cite{salzmann1998atomic,kluth,holladay2020accelerated}. The following processes are typically included in NLTE calculations:

\begin{itemize}
    \item electron-impact excitation and de-excitation
    \item electron-impact ionization and three-body recombination 
    \item photo-excitation and de-excitation 
    \item photo-ionization and radiative recombination 
    \item Brehmsstrahlung and inverse-Brehmsstrahlung 
    \item scattering
\end{itemize}

The above list is not exhaustive. There are other processes that occur at very high energies (greater than 1 MeV) as well as nuclear conversion processes that affect ionization levels and electron populations \cite{holladay1}.

LTE uses Kirchhoff's Law to assume that absorptivity and emissivity are equal and is only characterized by temperature; this makes look-up tables straightforward to implement. In contrast to LTE physics, temperature, density, and the radiative field characterize NLTE physics processes. In addition, the forward and backward processes are not equal in rate. This means that the counting of states and the subsequent solving of rate equations must be performed.

In terms of ICF simulations, two software packages are used in this paper, \hydra{} and \cretin{} \cite{kluth}. \hydra{} is a multiphysics package in which the overall simulation is performed \cite{hydra}. \cretin{} is the NLTE atomic physics package that calculates absorptivity and emissivity of elements which \hydra{} needs to perform the radiative transport calculations as part of the ICF simulation \cite{cretin}. \cretin{} often consumes seventy percent or more of the total computation time needed for ICF simulations. As a result, the fidelity and/or size of the simulations must be carefully balanced so as to fit within the time and resource constraints of the computer.

This is where machine learning, or more specifically neural networks, once again appears to be a great tool for improving performance via their simple evaluation and potential for  parallel scalability. Past work has shown that replacing the \cretin{} models with a set of neural networks, consisting of a fully-connected autoencoder and a deep jointly informed neural network (DJINN) model, can produce significant (10x) speed-up in the atomic physics for a single element when used in conjunction with \hydra{} \cite{kluth}. This use of an autoencoder and DJINN is analogous to reduced order modeling. The autoencoder, in this case, acts as a non-linear mapping to and from a low-dimensional space, and DJINN acts as a non-linear reduced order operator or solver on the low-dimensional space\cite{lee2020}.

\begin{figure}[H]
    \centering
    \begin{subfigure}[h]{.6\linewidth}
        \centering
        \includegraphics[width=\linewidth]{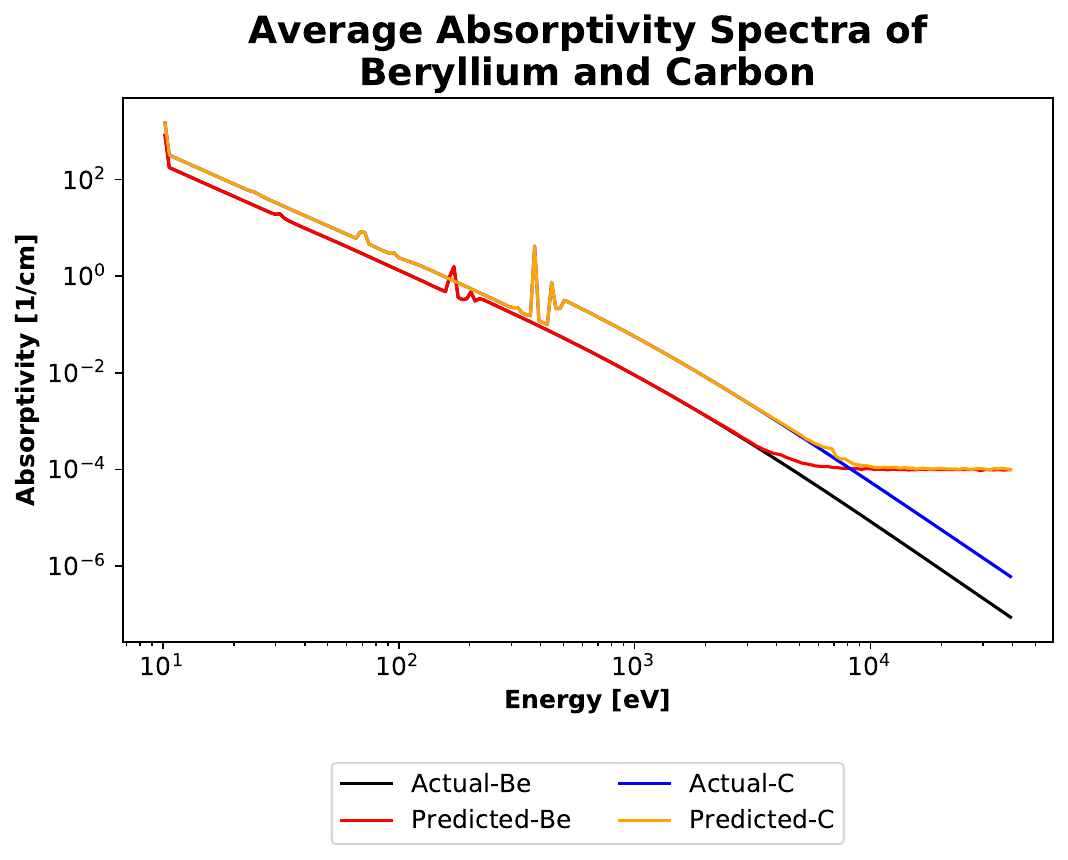}
    \end{subfigure}
    \caption{The average actual and predicted absorptivity spectra of beryllium and carbon.}
    \label{fig:problematic_spectra}
\end{figure}

This paper demonstrates the extension of results of the \cretin{} surrogate model to elements other than krypton, particularly for the low-atomic-number-element spectra. Those spectra are poorly reproduced with the proof-of-concept method as seen in Figure \ref{fig:problematic_spectra} where values below an apparent threshold are unable to be reproduce results for other elements that are similar to the results for krypton in \cite{kluth}. In addition, this paper is concerned with the ability to include multiple elements in a single neural network model for absorptivity or emissivity while also avoiding the threshold problem. As an important note the reader, this paper will focus largely on the autoencoders and producing a multi-element autoencoder. DJINN, for the most part, performs quite well at predicting the latent space of the autoencoder, thus the overall performance of the \cretin{} surrogate model is primarily dependent on the autoencoders' performance. Lastly, the goal of this paper is not to demonstrate better prediction performance of the multi-element model compared to individual models. Rather, the goal is to demonstrate that similar performance can be obtained.

\subsection{Related Work}
This paper builds upon the work of Kluth, et al. \cite{kluth}. The previous work utilized a fully-connected autoencoder which was trained to compress and decompress the spectra of krypton. Once trained, a DJINN model was trained to reproduce the latent space of the autoencoder using effectively the same input as \cretin{}: density, electron temperature, and a radiative field representation \cite{djinn,kluth}. The model as used in the implementation with \hydra{} used an autoencoder to produce a compressed representation of the radiative field calculated by \hydra{}. The radiative field for the surrogate model used a weighted sum of a Planckian and a Gaussian distribution at a given radiation temperature and a specified m-band ratio. The weighting factor is the m-band ratio ($\alpha)$ where the m-band is a Gaussian distribution centered at 3 keV with a full width half maximum of 1 keV \cite{kluth}.
As implemented with \hydra{}, the neural network model was able to provide a speed-up of 10x over \cretin{} for just the atomic physics and a 7x speed-up for the overall simulation. This speed up was accompanied by less than one percent relative error for both the electron temperature and radiative temperature as compared to when \hydra{} uses data directly from \cretin{} \cite{kluth}.


\section{Data}
\begin{figure}[H]
    \centering
        \includegraphics[width=.6\linewidth]{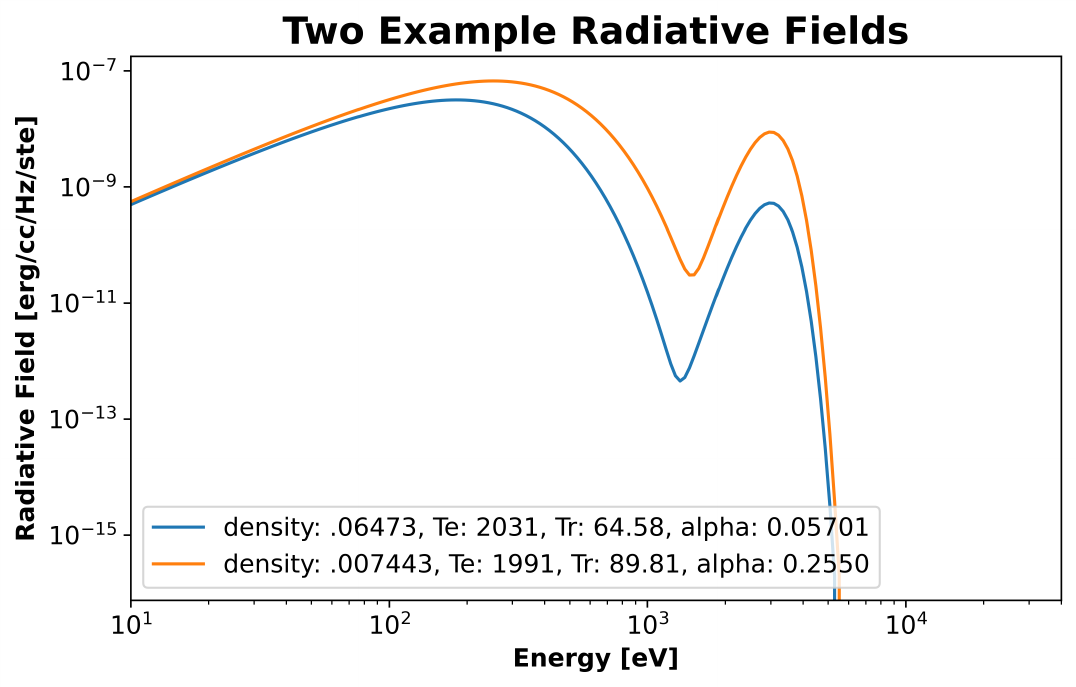}
    \caption{These are examples of two different radiative fields generated by the state process. The labels define the parameters used to generate the given radiative fields.}
    \label{fig:rad fields}
\end{figure}

The data that was used to train the neural networks was generated by repeated execution of \cretin{} with different input parameters using steady state approximations. The inputs were uniformly sampled values for density, electron temperature, radiation temperature, and the m-band ratio. The m-band is a 2 keV - 4 keV emissivity band that can roughly approximate the radiative fields expected in an ICF experiment \cite{li2011}. The absorptivity and emissivity are defined by the density, electron temperature, and the radiative field. An analytic expression is used to describe the radiative field, defined by radiative temperature and the m-band ratio using the function:
\begin{equation}
I(\nu) = aT^{4}_r \left[\left( 1-\alpha\right)b\left(\nu,T_r\right) \, + \, \alpha g\left(\nu \right) \right],    
\end{equation}
where, $a$ is the radiative constant of $7.5657*10^{-15}$ $\text{erg}/\text{cm}^3/\text{K}^4$, $b(\nu,T_r)$ is the reduced Planckian, and $g(\nu)$ is the previously mentioned Gaussian distribution centered at 3 keV with a full width half maximum set to 1 keV, and $\alpha$ is the m-band ratio \cite{kluth,li2011}. This is the same as how data for krypton was generated in \cite{kluth}. The ranges from which density, electron temperature, radiation temperature, and m-band ratio were sampled from were 0.003 g/cc to 0.1 g/cc, 300 eV to 3000 eV, 30 eV t o 300 eV, and 0.0 to 0.3 respectively. Figure \ref{fig:rad fields} shows two radiative fields generated in this manner. In total, 320,000 absorptivity spectra and 320,000 emissivity spectra with 200 logarithmically spaced energy bins are generated for beryllium, carbon, aluminum, iron, germanium, and krypton each. For clarification, one absorptivity spectra and one emissivity spectra are generated for each set of inputs.

The elements were selected based on the desire to have a large range of atomic numbers (z numbers). This allowed the exploration a wide variety of spectra shapes. The authors also wanted the elements to be as relevant to ICF experiments as possible, thus the authors chose beryllium, carbon, aluminum, iron, germanium, and krypton. 

Training and validation sets are created by giving each instance of a model its own 80/20 training/validation split, which is randomly generated. The eighty percent for training is further split 90/10 training/validation at training time for the purpose of having validation cost versus epoch statistics. The spectral data is scaled differently depending on which architecture is used for the autoencoder. The two forms of scaling used are a cube root transform and a $\log_{10}$ transform and are used either separately or in combination with each other. The $\log_{10}$ transform used the equation $x' = \log_{10}(x+1)$. Featurewise minmax scaling leads to rather poor performance. 

The data used to train the DJINN models consists of the randomly sampled values used to generate the absorptivity spectra and emissivity spectra as inputs as well as the z-numbers of the elements. The resultant latent space of the associated autoencoders is used as the output.


\section{Training}
One of the goals of this work is to demonstrate that similar performance to the proof-of-concept can be achieved for elements other than krypton while also exploring how the performance changes with respect to the atomic number of the elements.

\begin{table}[H]
    \centering
    \caption{Summary of the proof-of-concept architectures. $\log_{10}$ scaling was used for these models. The activation function used for all layers was the softplus function. Mean squared error was used as the cost function. \cite{kluth}}
    \label{tab:proof-of-concept arch}
    \begin{tabular}{rcc}
    \toprule
    &Absorption AE &Emission AE \\
    \midrule
    Input dimension     &200  &200 \\
    Latent space dimension &5 &7 \\
    Encoder &(96,46,22,10) &(124,77,48,29,18,11) \\
    Total Parameters: Encoder + Decoder &50349 &81287 \\
    \bottomrule
    \end{tabular}
\end{table}

\begin{figure}[H]
    \centering
    \includegraphics[height=1.5in]{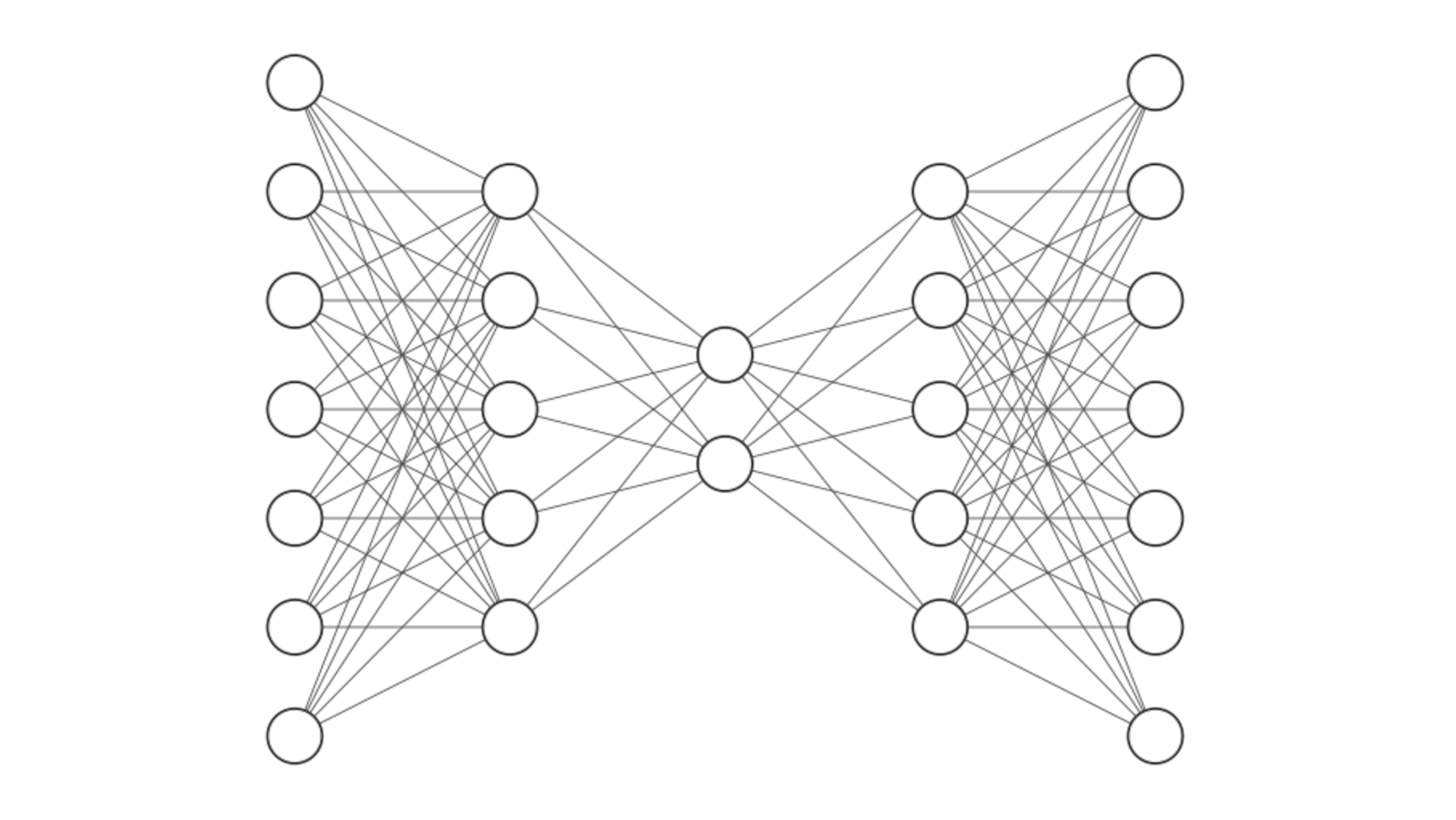}
    \caption{This is an example of a fully-connected autoencoder provided to demonstrate the hourglass shape seen in the majority of autoencoders. The latent space tends to be the smallest layer in the network. The left-hand side is the encoder, and the right-hand side is the decoder.}
    \label{fig:autoencoder}
\end{figure}

The architectures used in \cite{kluth} are different depending on whether the autoencoder is trained on absorptivity spectra or emissivity spectra with them both being fully-connected autoencoders. The absorptivity architecture consist of four hidden layers between the input layer and latent space layer that have widths which are harmonically decaying down to the latent space of five. This layer-size pattern is then mirrored over the latent space layer to complete the autoencoder. In general the shape of autoencoders tend to resemble the shape of the network in Figure \ref{fig:autoencoder}. The emissivity architecture consists of six hidden layers between the input layer and latent space layers that have widths which are harmonically decaying down to the latent space of seven. Both architectures used the softplus function as the activation function for all layers. \cite{kluth}. Table\ref{tab:proof-of-concept arch} summarizes each of these architectures. Both of the architectures are trained using $\log_{10}$ scaled spectra using a learning rate of 0.001 for 10,000 epochs with batch sizes consisting of 0.5\% of the total training data set (the 80\% fraction of the 320,000 spectra before the 90/10 split is applied). Adam is used as the optimizer, and mean square error is used as the cost function. The number of epochs, batch size, optimizer, and cost function stated here are used for all autoencoder models used in this paper.

\begin{table}[H]
    \centering
    \caption{Summary of the new single-element architecture. Cube root scaling was used for this model. The activation function used for all layers was the softplus function. Mean squared error was used as the cost function.}
    \label{tab:improved arch}
    \begin{tabular}{rc}
    \toprule
    &Absorption and Emission AE\\
    \midrule
    Input dimension     &200\\
    Latent space dimension &10\\
    Encoder &(100,33) \\
    Total Parameters: Encoder + Decoder &47736 \\
    \bottomrule
    \end{tabular}
\end{table}

A secondary goal of this work is to determine whether a single, fully-connected architecture could be found that performs as well or better than the proof-of-concept architectures while being smaller (use fewer parameters). The architecture selected is found by using various layer width scalings such as division by constant factors (2,3,4,...) or factorial division (2 then 3 then 4...). The latent space is experimented with separately from the layer width scaling. Summarized in table \ref{tab:improved arch}, new fully-connected architecture consists of only two hidden layers of widths 100 and 33 and a latent space dimension of 10. Like the proof-of-concept architectures, the activation function used is the softplus function. The learning rate is kept the same as what is used in the proof-of-concept. However, the scaling function is changed to a cube root scaling function for reasons that will be explained in the results section.

\begin{table}[H]
    \centering
    \caption{Summary of the multi-element architecture.}
    \label{tab:multi-element}
    \begin{tabular}{rc}
    \toprule
    &Absorption and Emission AE\\
    \midrule
    Input dimension     &200\\
    Latent space dimension &10\\
    Encoder: & \\
    Convolutional layer &1x5x20 \\
    & stride: 1 \\
    FC layers &(100,50,30) \\
    Total Parameters &818191 \\
    Encoder + Decoder & \\
    \bottomrule
    \end{tabular}
\end{table}

The third and primary goal is to determine whether multiple elements can be placed in a single model. Early experiments demonstrate that a fully-connected model would not be particularly effective. As a result, a convolutional architecture is adopted, which is summarized in table \ref{tab:multi-element}. The architecture consists of a convolutional layer that uses twenty 1-by-5 filters which are applied with a stride of one. The convolutional layer is followed by fully-connected layers of widths of 100, 50, 30 with a latent space dimension of 10. The activation function used for all layers is the exponential linear unit (ELU). It should be noted that, in order to keep training time below twenty-four hours, only forty percent of the full data set of each element is used in the convolutional network's data set. More explicitly, this means that 128,000 spectra came from each element as opposed to 320,000 spectra for a total of 768,000 spectra.

\begin{figure}[H]
    \centering
        \includegraphics[width=\linewidth]{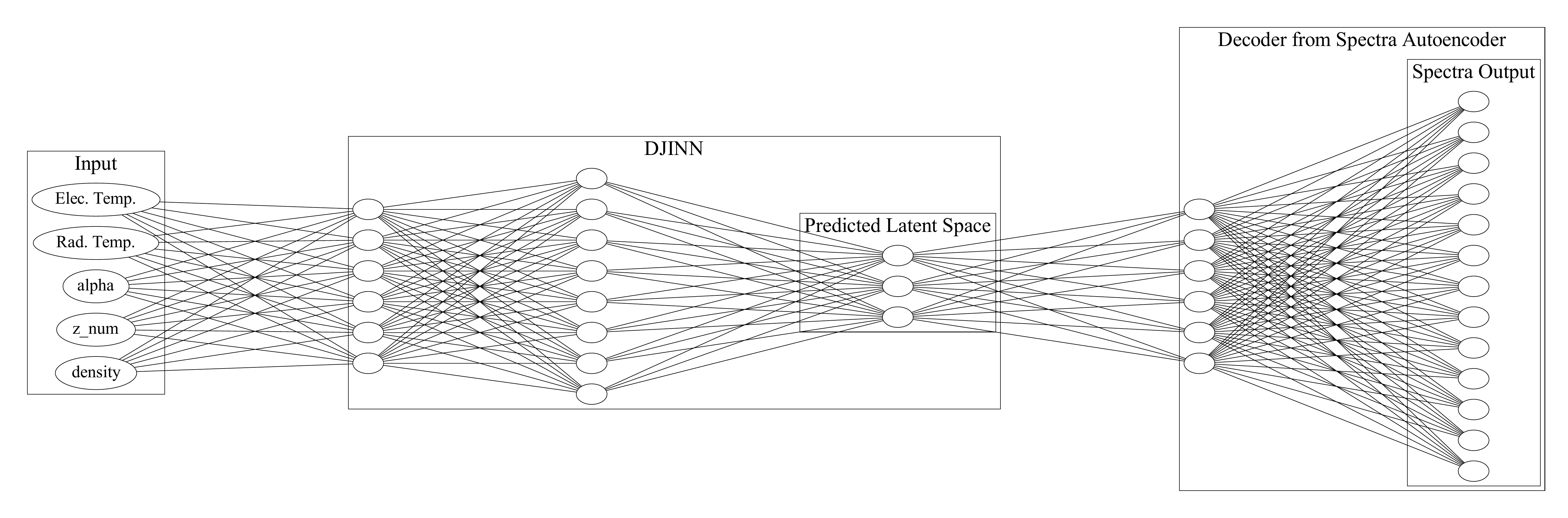}
    \caption{This neural network example graph demonstrates how the two different neural networks, DJINN and the decoder, are linked together. As used in this paper, these function as two individual networks that are trained separately.}
    \label{fig:djinn_flow}
\end{figure}

To verify the viability of a multi-element model beyond that of just an autoencoder, a DJINN model is trained to reproduce the latent spaces of each of the autoencoders. Figure \ref{fig:djinn_flow} shows the general process of how the \cretin{} surrogate model is put together. The DJINN model uses densities, electron temperatures, radiation temperatures, m-band ratios, and z-numbers as the inputs, and it is trained to reproduce the latent space of the autoencoder, the decoder is then used to map the latent space back to the full spectra. For DJINN, in all cases, the initial learning rate is set to 0.001, the number of trees is 1, the maximum depth is 11, and drop-out is turned off. The DJINN models are trained for 1000 epochs with a batch size of 0.5\% of the full size of the training data set.

\section{Results}
\subsection{Spectral Performance of Autoencoders}
\begin{figure}[H]
    \centering
    \begin{subfigure}[h]{.48\linewidth}
        \centering
        \includegraphics[width=\linewidth]{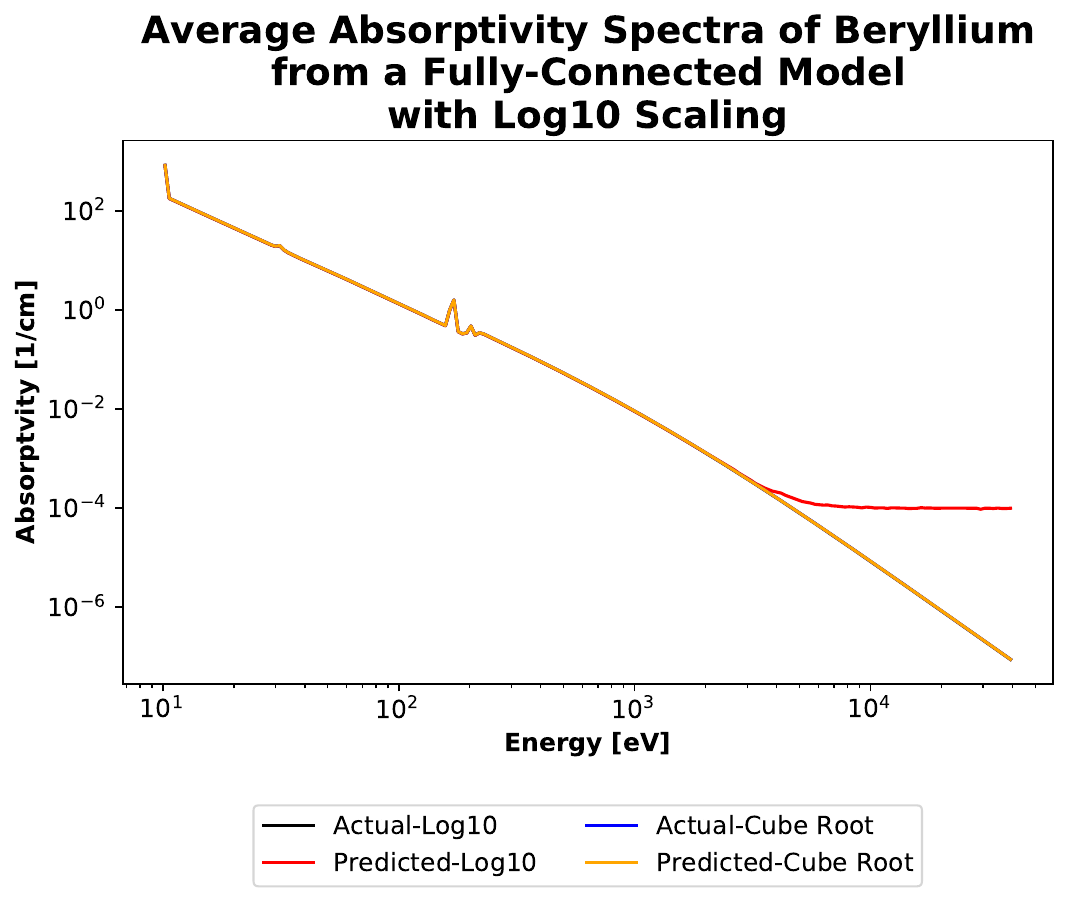}
    \end{subfigure}
    \begin{subfigure}[h]{.48\linewidth}
        \centering
        \includegraphics[width=\linewidth]{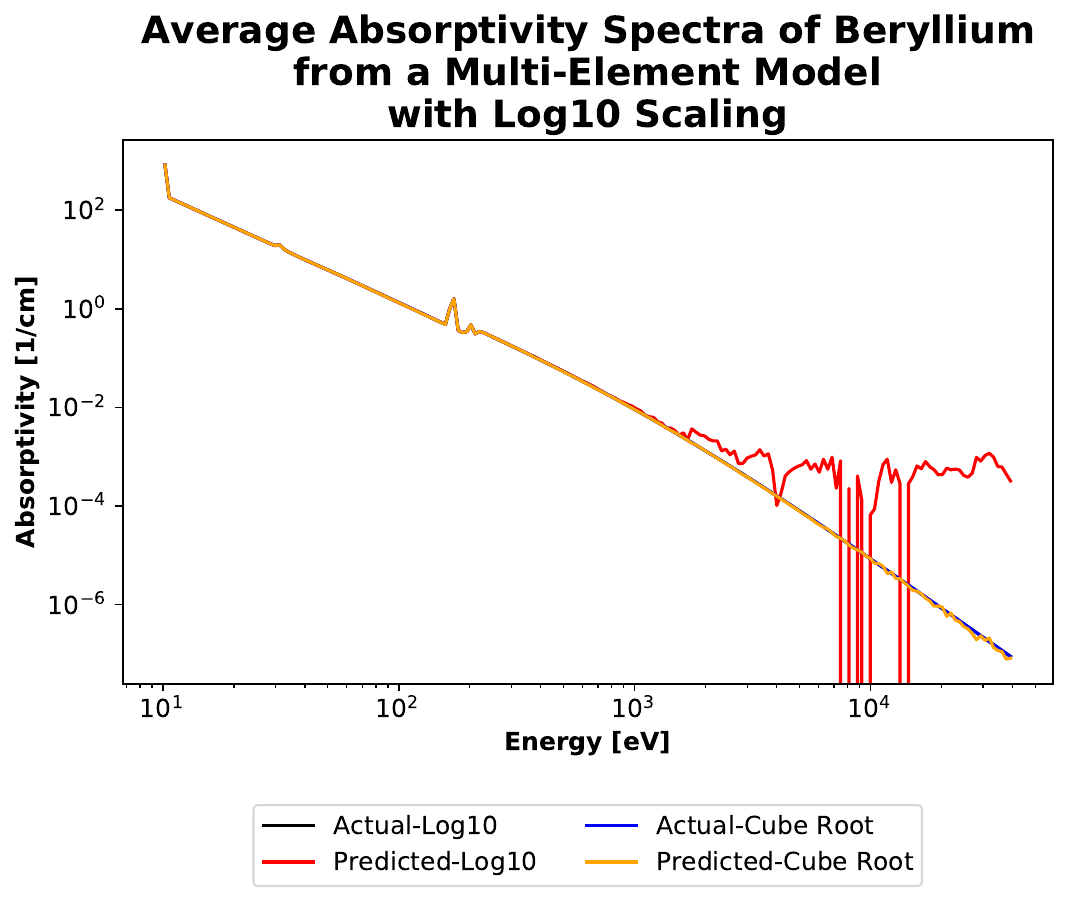}
    \end{subfigure}
    \caption{The issue that caused beryllium to have large error in its reproduction can occur in both a fully connected model or a convolutional model with multiple elements for both $\log_{10}$ and cube root scaling. It can be seen here that once the values reach a certain small value the network ceased to reproduce them properly. This problem does not manifest with the cube root scaling.}
    \label{fig:problem spectra}
\end{figure}

The plots in Figure \ref{fig:problem spectra}, address the immediate problem of the thresholding of predictions of the high-energy end of the spectra for the low-z elements. Both convolutional and fully-connected models are affected by the thresholding by the issue. The ``compression'' or reduction of the logarithmic magnitude of the data appears to be the solution to this issue.  It is interesting that $\log_{10}$ scaling does not the have the same effect. The small difference between, for example $log_{10}(1+.0001)$ and of $log_{10}(1+.00011)$. The absolute value of that difference is approximately $-4.34*10^{-6}$, and the evaluated values of those two scaled values are on the order of $1*10^{-5}$. That level order therefore places those values below the apparent threshold existing around $1*10^{-3}$.

Cube root scaling divides the logarithmic magnitude by three, thereby it effectively compresses the values as small as $1*10^{-9}$ into a usable range. The effect is not completely equal for both the fully-connected models and convolutional models. Though the ripple is small, one does exist in the convolutional model's capacity to predict the smaller values of the spectra. In contrast, the fully-connected model produces results that are indistinguishable in the Figure \ref{fig:problem spectra}.

\begin{figure}[H]
    \centering
    \begin{subfigure}[h]{.48\linewidth}
        \centering
        \includegraphics[width=\linewidth]{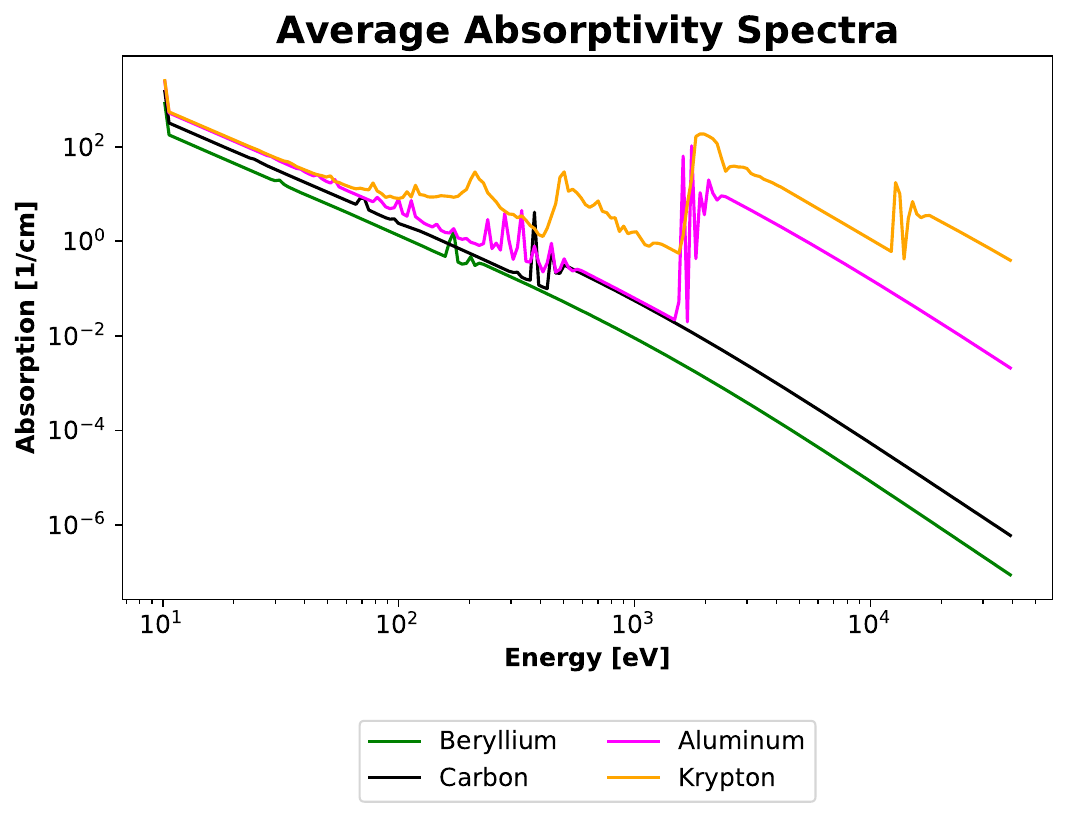}
    \end{subfigure}
    \begin{subfigure}[h]{.48\linewidth}
        \centering
        \includegraphics[width=\linewidth]{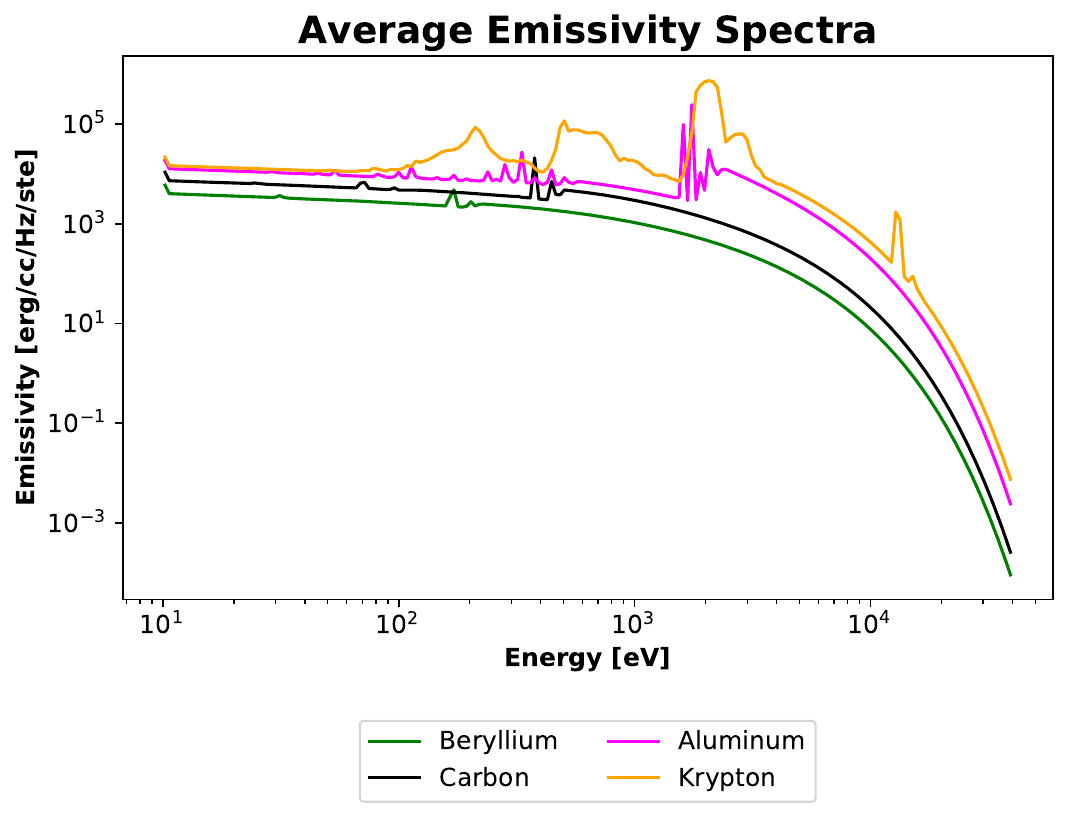}
    \end{subfigure}
    \caption{These are the average absorptivity and emissivity spectra for selected elements. Iron and germanium were not included to improve the readability of the plots.}
    \label{fig:average spectra}
\end{figure}

As to why a beryllium and carbon exhibit this thresholding issue and why it did not seem to appear it in the proof-of-concept work \cite{kluth}, it is because as the z-number of elements increases, the mean logarithmic magnitude of the spectra also increases. Figure \ref{fig:average spectra} shows this trend. This difference in logarithmic magnitude largely arises from the large spikes in absorptivity for the high-z elements which do not exist in the low-z elements. Spectra also exhibit a general upward shift in values which can be seen on the low-energy end of the spectra. It should be noted, however, the issue of thresholding does still exist in the high-z elements. It is just much less noticeable and only appears for low-temperature and low-density inputs.

\begin{figure}[H]
    \centering
    \begin{subfigure}[h]{.48\linewidth}
        \centering
        \includegraphics[width=\linewidth]{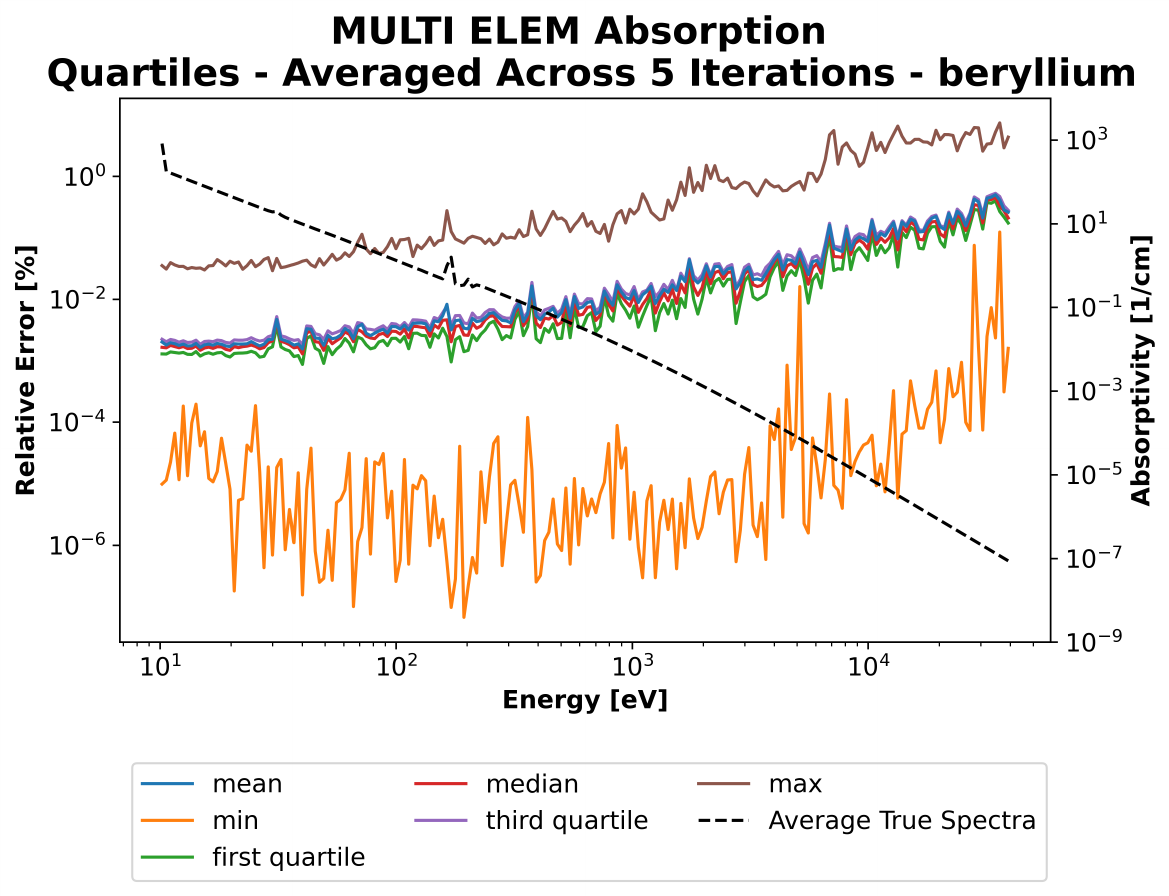}
    \end{subfigure}
    \begin{subfigure}[h]{.48\linewidth}
        \centering
        \includegraphics[width=\linewidth]{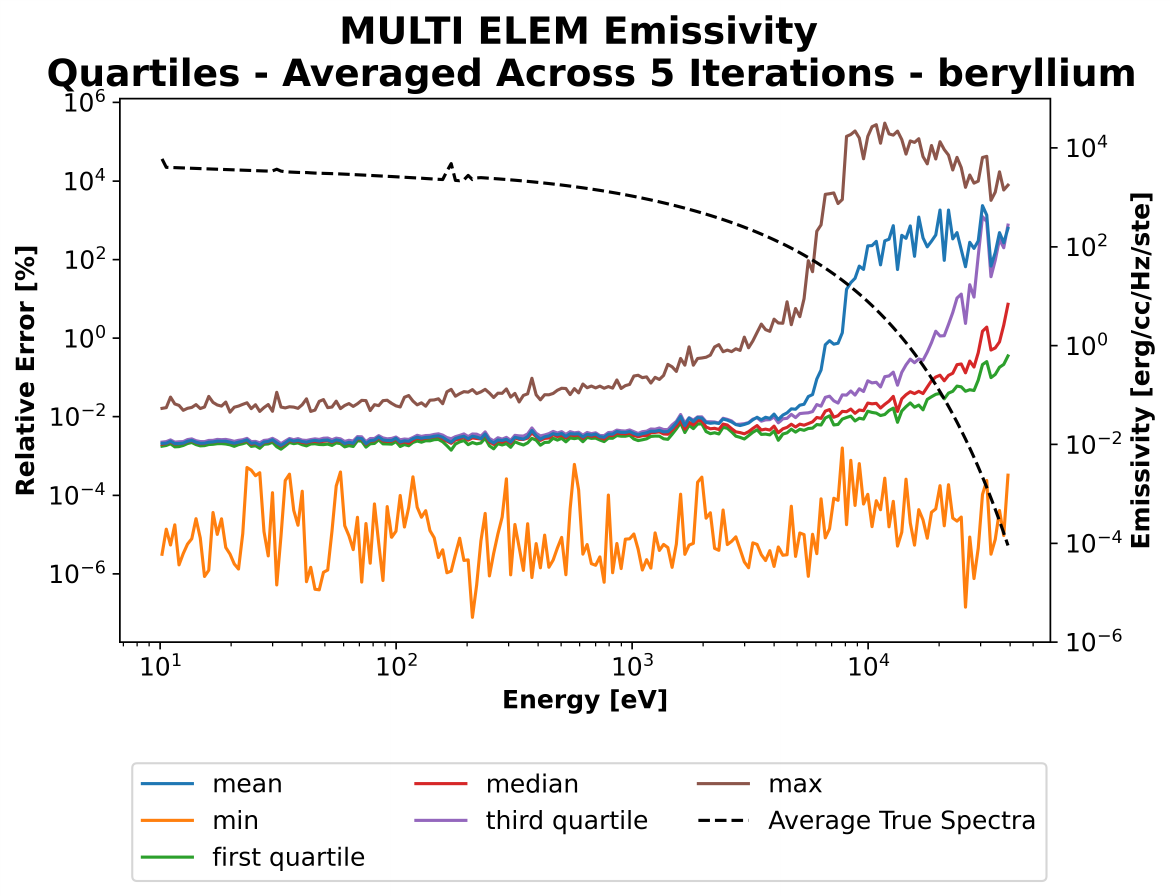}
    \end{subfigure}
    \caption{These plots show for the absorptivity and emissivity of beryllium the minimum, mean, maximum, quartiles, and the average spectra of the beryllium samples used. In the left plot it can be seen that the maximum error for absorptivity does exceed 1\%. Also seen is that the middle quartiles are tightly clustered, but along with maximum error, the errors increase in a rather linear fashion as the energy goes up. The emissivity of beryllium, in the right plot, sees a large increase in maximum relative error around 10 keV which also results in a large increase in mean relative error. The middle quartiles do begin to diverge from each other around 20 keV.}
    \label{fig:multi elem be quartiles}
\end{figure}

\begin{figure}[H]
    \centering
    \begin{subfigure}[h]{.48\linewidth}
        \centering
        \includegraphics[width=\linewidth]{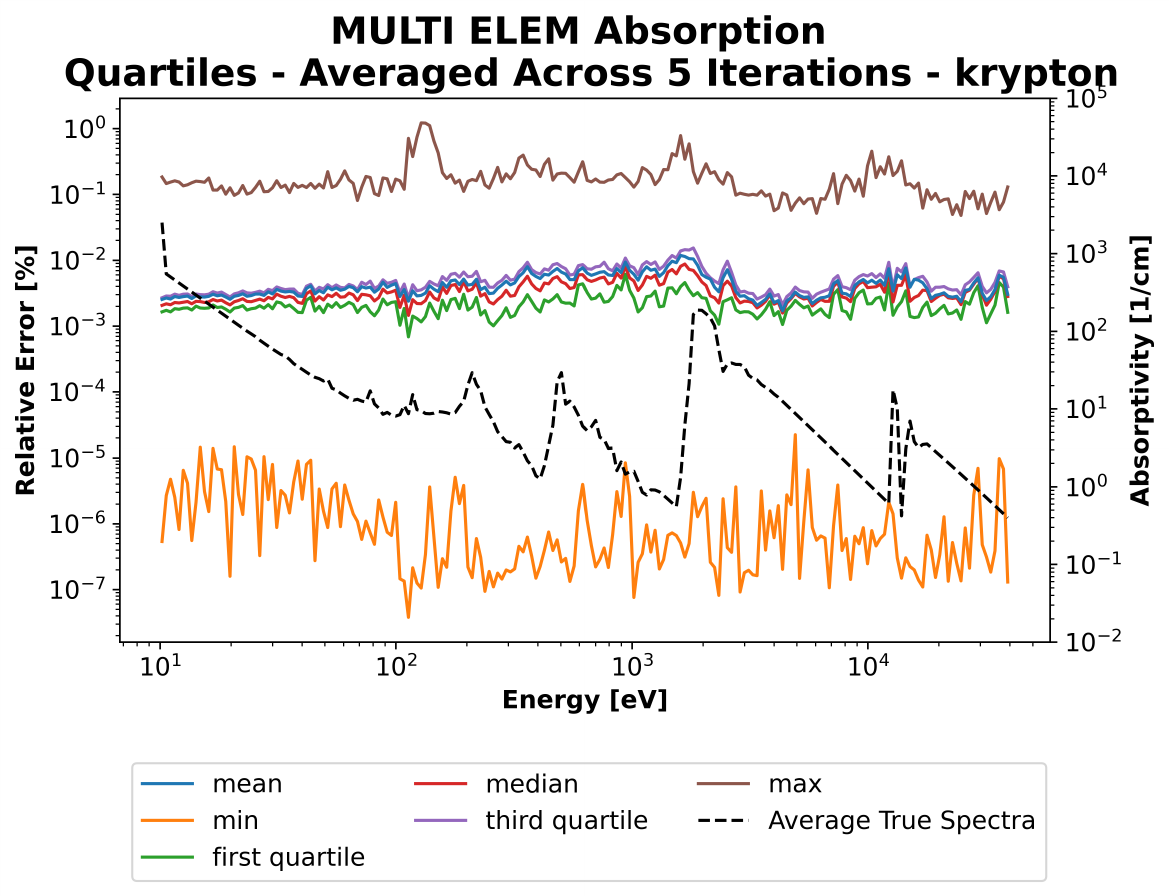}
    \end{subfigure}
    \begin{subfigure}[h]{.48\linewidth}
        \centering
        \includegraphics[width=\linewidth]{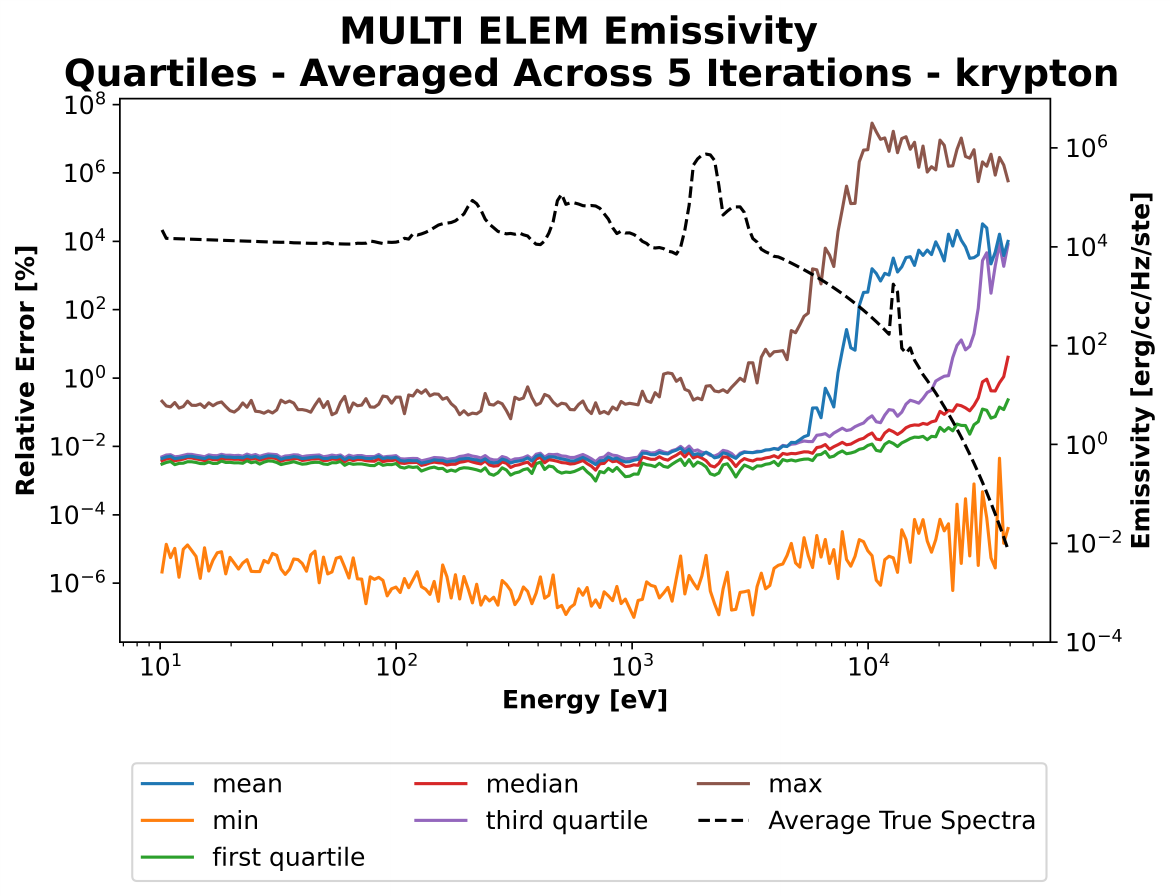}
    \end{subfigure}
    \caption{These plots show for the absorptivity and emissivity of krypton the minimum, mean, maximum, quartiles, and the average spectra of the krypton samples used. In the left plot it can be seen that the maximum error for absorptivity may just barely exceed 1\%, but the performance is different from that of beryllium in that the maximum relative error is fairly stable across the entire energy range. Also, seen are that the middle quartiles are in close proximity to each other. Much like beryllium, krypton sees a large increase in maximum error for emissivity around 10 keV with a corresponding increase in mean error. The middle quartiles begin to diverge in a way that appears nearly identical to that of beryllium.}
    \label{fig:multi elem kr quartiles}
\end{figure}

The plots in Figures \ref{fig:multi elem be quartiles} and \ref{fig:multi elem kr quartiles} consist of the minimums, first quartiles, medians, means, third quartiles, and maximums of the bin-wise percent relative error averaged over each of the five iterations of the tested model. For example, the first quartiles are the average of the first quartiles over the five models. The average of the true spectra is then plotted on top of these to provide an idea of what spectral features may lead to increases in error.
Thus, from Figures \ref{fig:multi elem be quartiles} and \ref{fig:multi elem kr quartiles}, a general idea of the uncertainty of the multi-element model can be obtained. Both the absorptivity and emissivity for both beryllium and krypton have a third-quartile that sticks very close to the median except at the highest of energy levels. Additionally, the third-quartiles of the emissivity errors stay below ten percent error until about 20 keV which is generally outside of the scope of ICF simulations. With the exception of the low-z elements the absorptivity spectra have a rather flat level of error. While this does imply that a full quarter of the spectra produced errors even higher, those spectra still perform well below 0.1\% relative error for the half or more of the full spectra. The equivalent plots for the single-element models, which are not shown here, possess increasing errors for the high energy portion of the spectra, but the maximum errors are not as large. The point where the error starts climbing in the single-element models is delayed compared to the multi-element model. However, an issue with the single-element models is that they appear to be more prone to produce increased errors around sharp features in the spectra. This is opposed to the multi-element model, a convolutional model, which is not immune but more resistant to such issues. The majority of this increasing error is thought to be contributed by the increasingly small values which, for emissivity, can be as low as $1*10^{-49}$. 


\subsection{Gray Approximation of Autoencoder}
Error results here are reported as percent relative error of a simple gray approximation of the spectra. A gray approximation is the summation or integral of either the absorptivity or emissivity over the energy range of the spectra. The two primary statistics presented are the mean relative error and the maximum relative error of these gray approximations. The mean relative errors as reported are the arithmetic means of the mean percent relative error of five different models trained with the same architecture. The maximum relative error is the maximum percent relative error for any given spectra from any of five models having the same architecture. The errors are also reported after re-scaling back to the physical dimensions.

\begin{figure}[H]
    \centering
    \begin{subfigure}[h]{.48\linewidth}
        \centering
        \includegraphics[width=\linewidth]{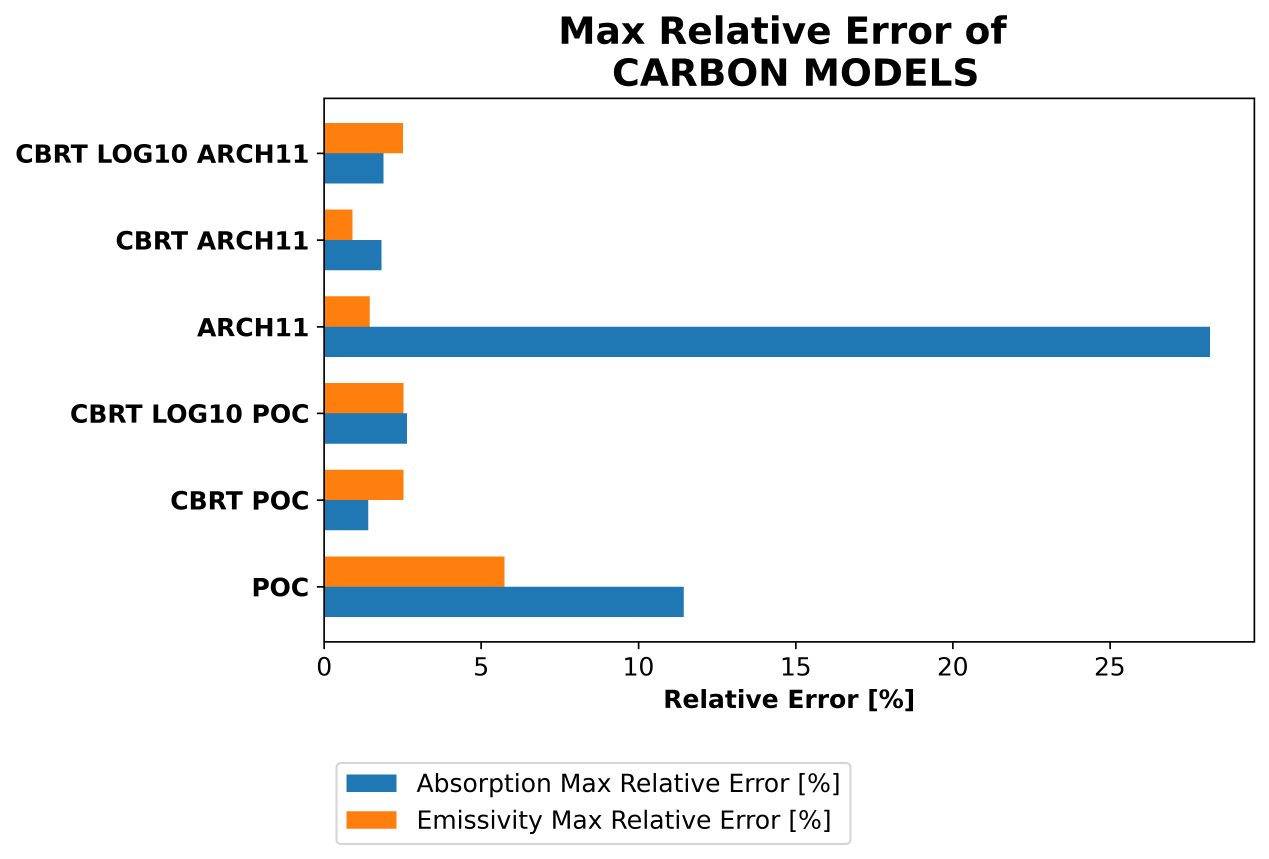}
    \end{subfigure}
    \begin{subfigure}[h]{.48\linewidth}
        \centering
        \includegraphics[width=\linewidth]{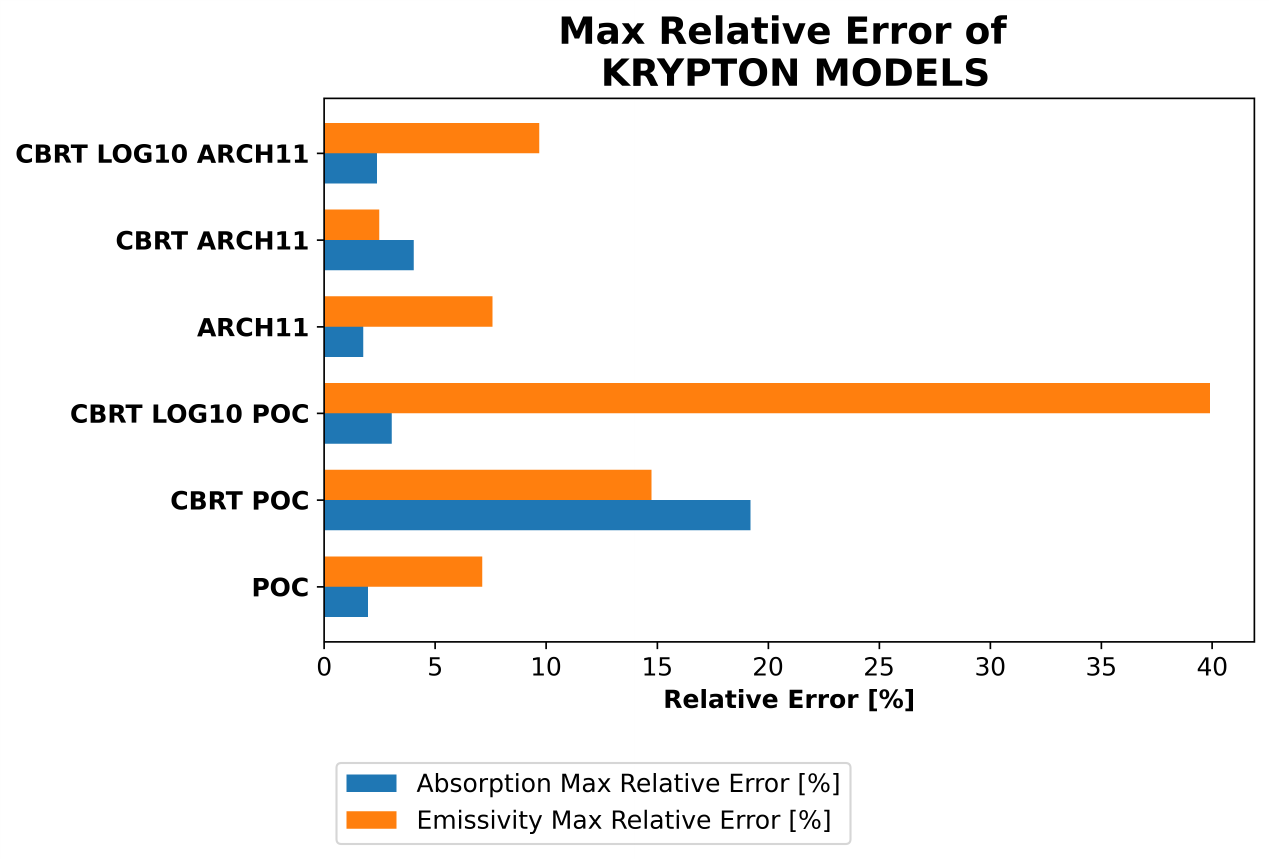}
    \end{subfigure}
    \caption{The maximum relative errors for the various single element models of carbon and krypton. `CBRT' denotes the use of cube root scaling. LOG10 refers to the use of $\log_{10}$ scaling. `CBRT LOG10' references the use of cube root scaling followed by $\log_{10}$ scaling. `POC' denotes the proof-of-concept architecture, and `ARCH11' denotes that the architecture selected was the eleventh architecture tested.}
    \label{fig:arch11 variations}
\end{figure}

 Figure \ref{fig:arch11 variations} is provided to offer a look at what happens to the maximum relative error with the change of scaling functions. Here it can be seen that the maximum error for the $\log_{10}$ scaled models explode for the absorptivity of carbon. The new fully-connected krypton model does not have a better maximum error for absorptivity, but it does have better maximum error emissivity. The other elements actually displayed higher tendencies to perform better on the maximum error with the new fully-connected architecture, but the mean errors are a toss up between the cube root scaled new architecture and the $\log_{10}$ scaled proof-of-concept architecture. The performance on the low-z elements ultimately decides which architecture is better. The authors note that hyperparameter tuning has been explored, but changing parameters such as the learning rate produce only slightly better mean relative errors with slightly worse maximum errors. This behavior is not entirely surprising because lower learning rates, for example, are often associated with worse generalization which often translates to higher maximum relative errors and sometimes lower mean relative errors \cite{liNeurIPS2019}.

\begin{table}[t]
    \centering
    \caption{This table contains the mean percent relative error and maximum percent relative error of the multi-element model as well as the new fully-connected (FC) model and the proof-of-concept (POC) model. The multi-element model consisted of iterations of the same model. The new FC model and the POC model results are the results of the individual single-element models.}
    \label{tab: multi-elem}
    \begin{tabular}{rrccccccc}
        \toprule
        \multicolumn{2}{c}{Absorptivity}     & Be & C & Al & Fe & Ge & Kr & All \\
        \midrule
        \multirow{2}{*}{Multi-element} & mean & .181  & .208 & .231     & .165   & .176   & .196    & .211     \\  
                                       & max  & 5.02  & 4.87 & 3.59     & 2.64   & 7.32   & 13.4    & 13.4    \\
        \midrule
        \multirow{2}{*}{New FC}   & mean & .056   & .365  & .068    & .137   & .316   & .087    & -   \\ 
                                       & max  & .882   & 1.83  & 3.23    & 2.46   & 2.82   & 4.04    & -   \\
        \midrule
        \multirow{2}{*}{POC}           & mean & .912   & .473  & .155    & .094   & .151   & .149    & -   \\  
                                       & max  & 19.1   & 11.4  & 10.3    & 3.35   & 2.80   & 1.99    & -   \\
        \bottomrule
    \end{tabular}
    \begin{tabular}{rrccccccc}
        \toprule
        \multicolumn{2}{c}{Emissivity}     & Be & C & Al & Fe & Ge & Kr & All \\
        \midrule
        \multirow{2}{*}{Multi-element} & mean & .161  & .238 & .172    & .215   & .217   & .189    & .131    \\  
                                       & max  & 4.24  & 4.77 & 5.73    & 1.97   & 1.62   & 12.5    & 12.5    \\ 
        \midrule
        \multirow{2}{*}{New FC}   & mean & .167   & .056  & .119    & .123   & .109   & .182    & -   \\ 
                                       & max  & .672   & .905  & 5.87    & 2.42   & 3.10   & 2.49    & -   \\
        \midrule
        \multirow{2}{*}{POC}           & mean & .248   & .589  & 1.53    & .349   & .560   & .411    & -   \\  
                                       & max  & 3.12   & 5.74  & 50.0    & 11.3   & 15.3   & 7.13    & -   \\
        \bottomrule
    \end{tabular}
\end{table}

Attempts to create a multi-element model are reasonably successful in that the model achieved an overall mean relative error 0.211\% and 0.121\% for absorptivity and emissivity respectively, but the model is larger, in terms of the number of parameters, than would be ideal considering the model has the approximately equivalent memory footprint of seventeen, fully-connected models. Attempts to make smaller multi-element models have been unsuccessful due to large mean relative errors and an increase in training time.

Table \ref{tab: multi-elem} more fully shows that the multi-element model performs comparably to the two different single-element model architectures on mean relative error for absorptivity and even out performs the proof-of-concept mean error values for low-z elements. The multi-element model also performs better than the proof-of-concept architecture for emissivity in mean error for each of the elements; however, the new architecture tends to perform better for both absorptivity and emissivity on mean error. In terms of absorptivity maximum error, the multi-element model generally performs comparably with the new architecture and outperforms the proof-of-concept model on the four lightest elements. This is not completely mirrored by the emissivity maximum error.

\subsection{Latent-space predictions using DJINN}
\begin{figure}[h]
    \centering
    \begin{subfigure}[h]{.48\linewidth}
        \centering
        \includegraphics[width=\linewidth]{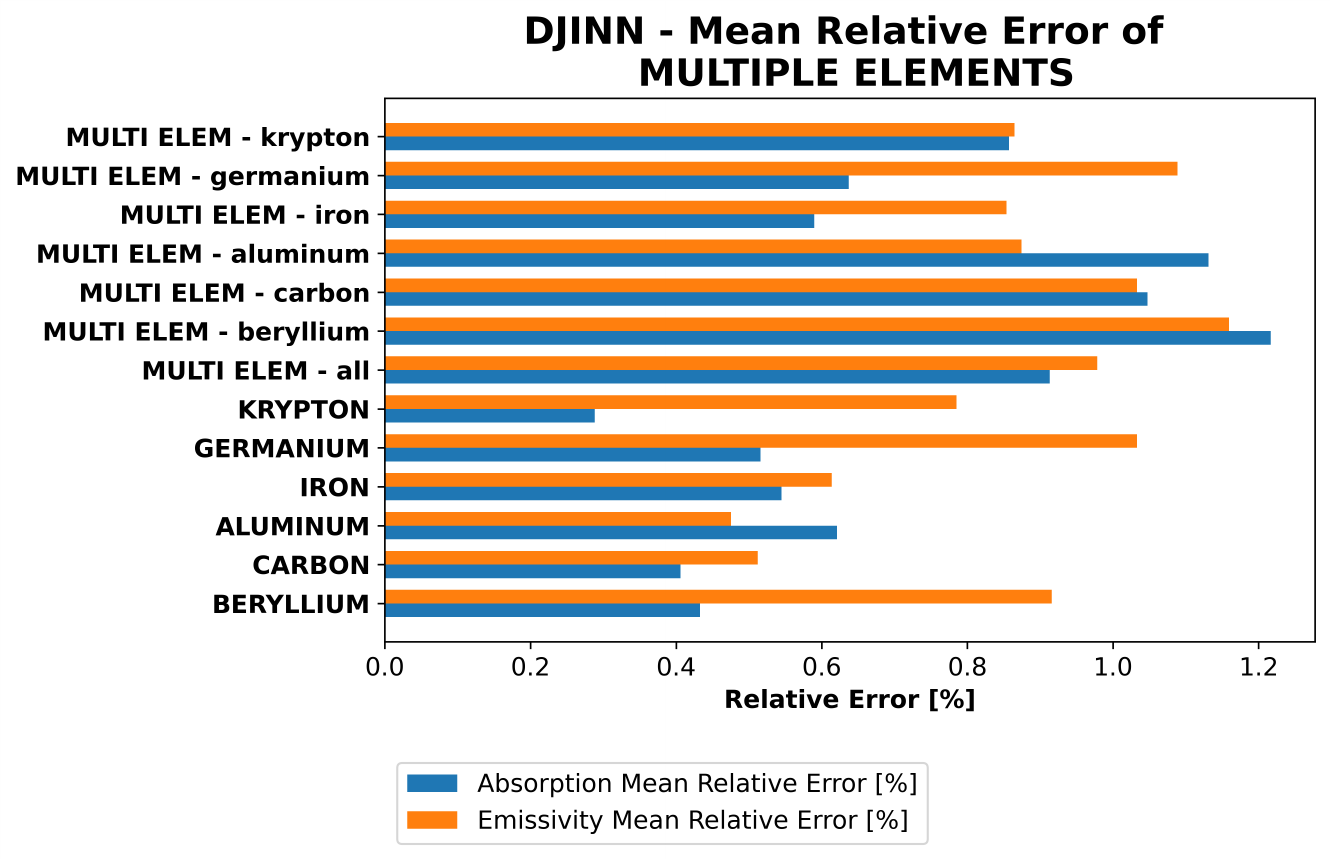}
    \end{subfigure}
    \begin{subfigure}[h]{.48\linewidth}
        \centering
        \includegraphics[width=\linewidth]{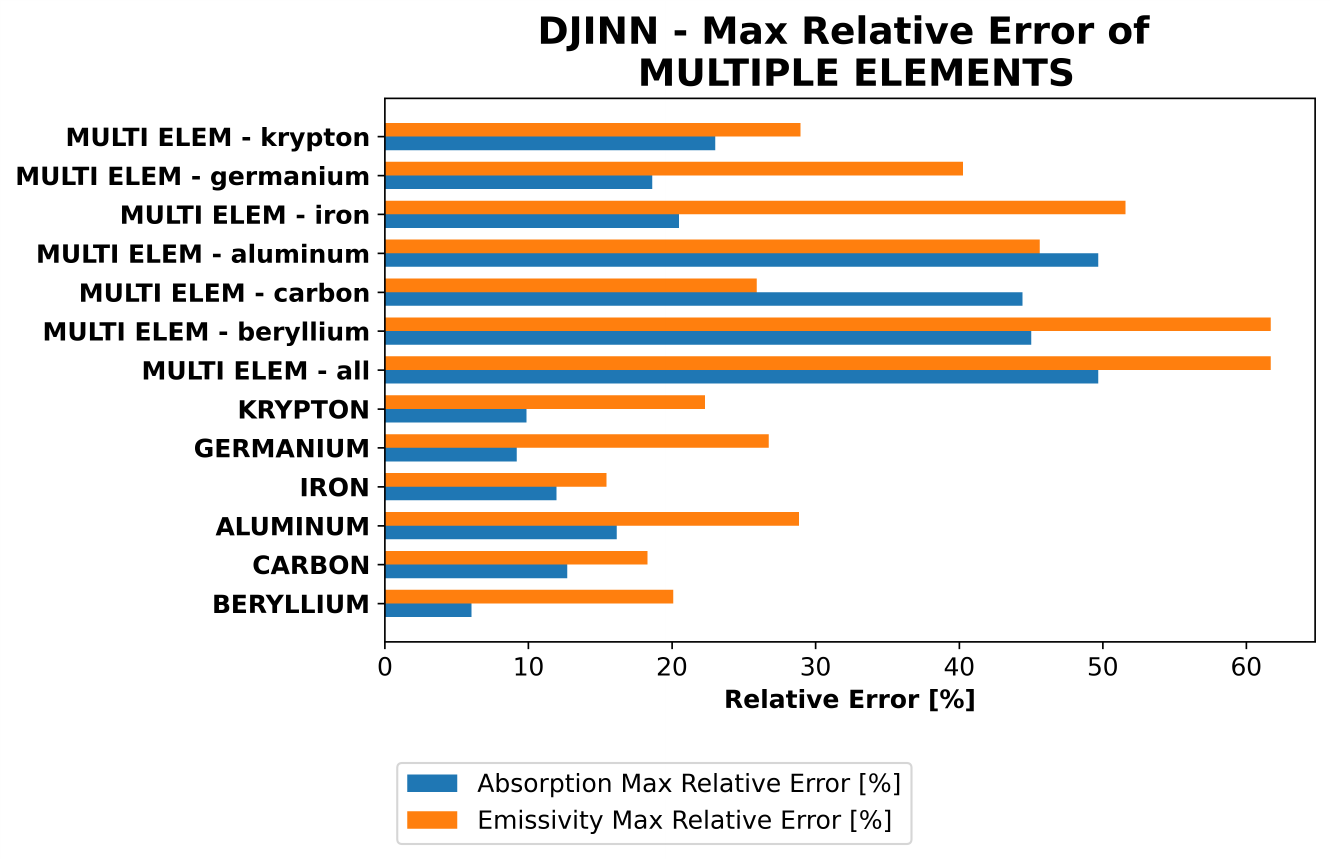}
    \end{subfigure}
    \caption{The maximum relative errors for the various single element models of carbon and krypton. All of these models used a cube root scaling.}
    \label{fig:djinn multiple elem}
\end{figure}

\begin{figure}[h]
    \centering
    \begin{subfigure}[h]{.48\linewidth}
        \centering
        \includegraphics[width=\linewidth]{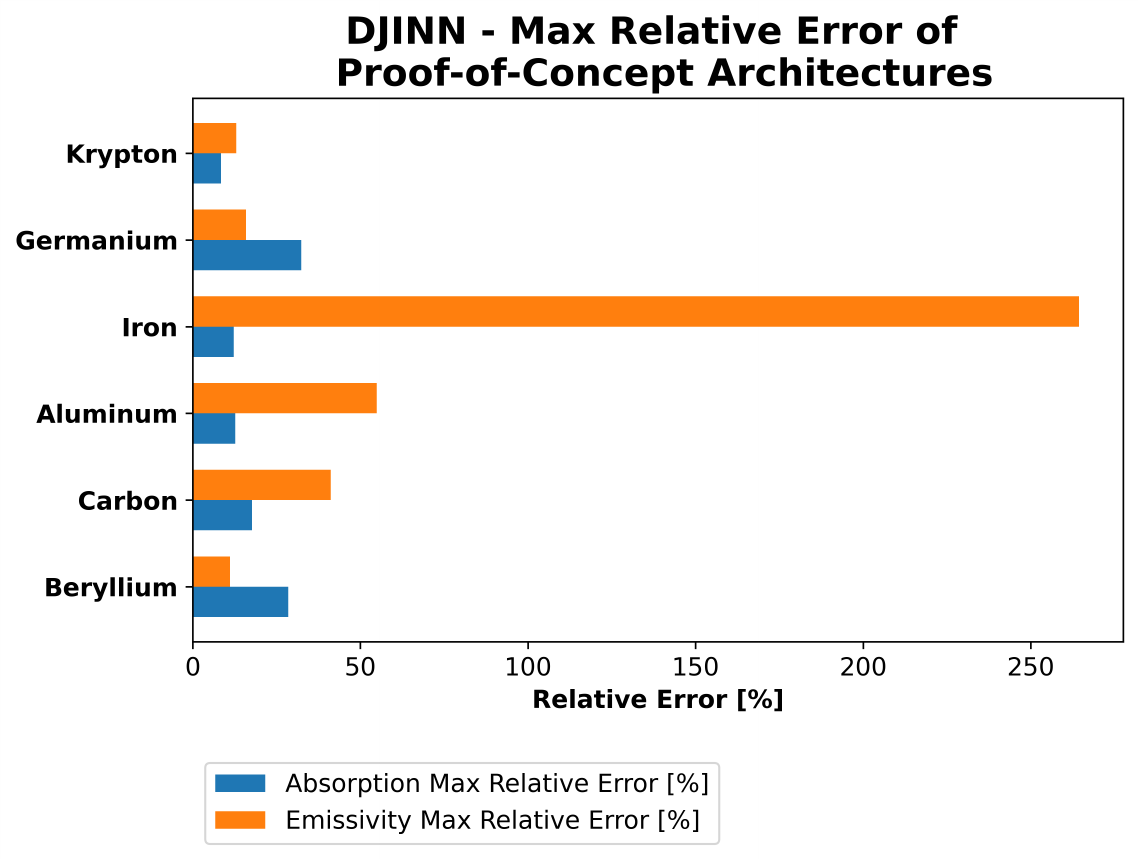}
    \end{subfigure}
    \begin{subfigure}[h]{.48\linewidth}
        \centering
        \includegraphics[width=\linewidth]{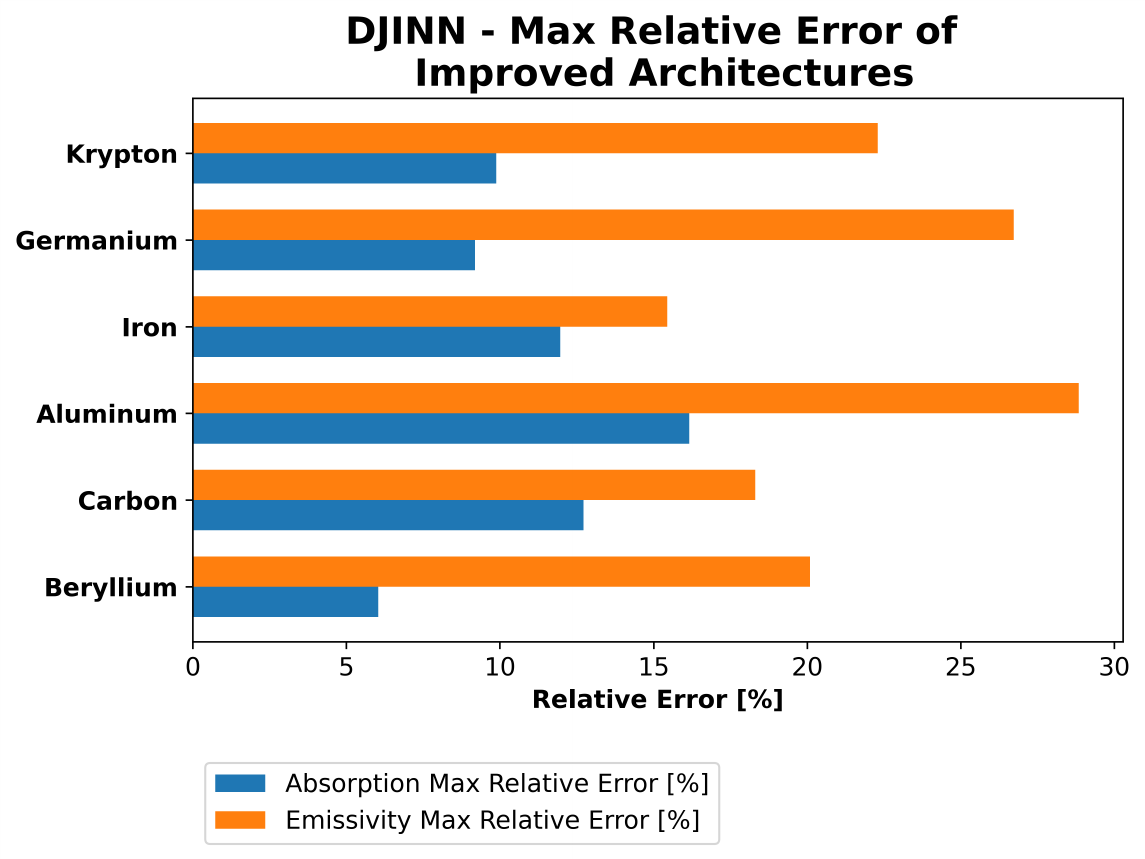}
    \end{subfigure}
    \caption{The maximum relative errors for the various single element models of carbon and krypton. `CBRT' denotes the use of cube root scaling. LOG10 refers to the use of $\log_{10}$ scaling. `CBRT LOG10' references the use of cube root scaling followed by $\log_{10}$ scaling. `POC' denotes the proof-of-concept architecture, and `ARCH11' denotes that the architecture selected was the eleventh architecture tested.}
    \label{fig:djinn poc v improved}
\end{figure}

After the selection of the new fully-connected architectures and multi-elements architectures, DJINN models are trained to see how the combined models compare to the performance of the proof-of-concept results. Figure \ref{fig:djinn multiple elem} shows that the mean relative error of the multi-element models performs generally pretty well and reasonably close to the single-element models. The maximum relative error, though, does not perform as well. The DJINN model introduces enough noise that the maximum relative errors is generally multiplied by a factor between 10 and 100.
Figure \ref{fig:djinn poc v improved} shows the maximum relative error of the single-element models. The emissivity of iron as reproduced with the proof-of-concept autoencoder plus DJINN does poorly, and it makes it difficult to tell that krypton and germanium tend to perform better with the model that uses the proof-of-concept autoencoder. The mean relative error results, which can be seen in Table \ref{tab:djinn multi-elem} for the low-z elements, point to the models that use the new fully-connected autoencoders being better. This generally holds for the maximum relative error, but the high-z elements tend to do a bit better with the models that use the proof-of-concept autoencoder. This is likely explained by the greater number parameters in the network being to represent the increased number of features in the high-z elements' spectra.

\begin{table}[t]
    \centering
    \caption{This table contains the mean percent relative error and maximum percent relative error of the combined multi-element autoencoder and its associated DJINN model  as well as the new fully-connected (FC) model and the proof-of-concept (POC) model in combinations with their respective DJINN model. The multi-element model results all come from one convolutional model plus its associated DJINN model. The new FC model and the POC model results are the those of the individual single-element models with each of them having its own DJINN model.}
    \label{tab:djinn multi-elem}
    \begin{tabular}{rrccccccc}
        \toprule
        \multicolumn{2}{c}{Absorptivity}     & Be & C & Al & Fe & Ge & Kr & All \\
        \midrule
        \multirow{2}{*}{Multi-element} & mean & 1.21  & 1.05 & 1.13     & .590   & .637   & .857    & .913     \\  
                                       & max  & 45.0  & 44.4 & 49.7     & 20.5   & 18.6   & 23.0    & 49.7    \\
        \midrule
        \multirow{2}{*}{New FC}   & mean & .433   & .406  & .621    & .545   & .516   & .288    & -   \\ 
                                       & max  & 6.04   & 12.7  & 16.2    & 12.0   & 9.19   & 9.88    & -   \\
        \midrule
        \multirow{2}{*}{POC}           & mean & .929   & 1.07  & .391    & .310   & .497   & .305    & -   \\  
                                       & max  & 28.5   & 17.7  & 12.7    & 12.2   & 32.4   & 8.42    & -   \\
        \bottomrule
    \end{tabular}
    \begin{tabular}{rrccccccc}
        \toprule
        \multicolumn{2}{c}{Emissivity}     & Be & C & Al & Fe & Ge & Kr & All \\
        \midrule
        \multirow{2}{*}{Multi-element} & mean & 1.16  & 1.03 & .875 & .854  & 1.09   & .865    & .978    \\  
                                       & max  & 61.7  & 25.9 & 45.6 & 51.6  & 40.3   & 29.0    & 61.7    \\ 
        \midrule
        \multirow{2}{*}{New FC}   & mean & .916   & .512  & .475   & .614   & 1.03   & .785    & -   \\ 
                                       & max  & 20.1   & 18.3  & 28.8   & 15.4   & 26.7   & 22.3    & -   \\
        \midrule
        \multirow{2}{*}{POC}           & mean & 1.03   & 1.72  & 1.77   & .741   & .806   & .791    & -   \\  
                                       & max  & 11.1   & 41.2  & 55.0   & 264    & 15.9   & 13.0    & -   \\
        \bottomrule
    \end{tabular}
\end{table}

\section{Conclusion and Future Work}
It has been shown here that the previous work done in the proof-of-concept \cite{kluth} can be reproduced for various other elements. 
This work demonstrates that low-z elements, or rather low values of absorptivity and emissivity, can be more accurately reproduced when using cube root scaling as opposed to the original $\log_{10}$ scaling. It is also likely that further compression with a higher order root may be even more beneficial. A convolutional autoencoder is found to be an effective solution to encode and decode the absorptivity and emissivity spectra of multiple elements in a single model with a similar level of accuracy as the single-element models. Lastly, it is demonstrated that DJINN, with the inclusion of the elements' z-numbers as inputs, can learn the latent space of the decoder of the multi-element model, but with significantly more uncertainty than the single element models. Further experimentation with the hyperparameters used when training the DJINN model may potentially improve these results.

Directly continuing from this work will include better tuning of the DJINN model, so the multi-element model may obtain performance far closer to that demonstrated by the single-element models. This process will also likely include the implementation of transfer learning in the sense of linking multiple pre-trained models end to end and proceeding to train it further. Ultimately, the results here will be used to improve the speed at which \hydra{} simulations can be performed by replacing \cretin{} for multiple element computations. 


\section*{Acknowledgements}
This work was performed under the auspices of the U.S. Department of Energy by Lawrence Livermore National Laboratory under Contract DE-AC52-07NA27344. Released as LLNL-JRNL-822718-DRAFT.

This document was prepared as an account of work sponsored by an agency of the United States government. Neither the United States government nor Lawrence Livermore National Security, LLC, nor any of their employees makes any warranty, expressed or implied, or assumes any legal liability or responsibility for the accuracy, completeness, or usefulness of any information, apparatus, product, or process disclosed, or represents that its use would not infringe privately owned rights. Reference herein to any specific commercial product, process, or service by trade name, trademark, manufacturer, or otherwise does not necessarily constitute or imply its endorsement, recommendation, or favoring by the United States government or Lawrence Livermore National Security, LLC. The views and opinions of authors expressed herein do not necessarily state or reflect those of the United States government or Lawrence Livermore National Security, LLC, and shall not be used for advertising or product endorsement purposes.

A thank you goes to Gilles Kluth for help in running \cretin{} as well as providing the krypton data.

\bibliographystyle{elsarticle-num}
\bibliography{cas-refs}





\end{document}